\documentclass[twocolumn,showpacs,preprintnumbers,amsmath,amssymb,floatfix]{revtex4}

\usepackage{graphicx}
\usepackage{dcolumn}
\usepackage{bm}

\newcommand{\pom}{\tt I\! P}
\newcommand{\beq}{\begin{equation}}
\newcommand{\eeq}{\end{equation}}

\begin{document}

\title{Hard diffractive quarkonium hadroproduction at high energies}

\author{M. V. T. Machado}

\affiliation{Centro de Ci\^encias Exatas e Tecnol\'ogicas, Universidade Federal do Pampa \\
Campus de Bag\'e, Rua Carlos Barbosa. CEP 96400-970. Bag\'e, RS, Brazil}

\begin{abstract}

We present a study of heavy quarkonium production in hard diffractive process by the Pomeron exchange for Tevatron and LHC energies. The numerical results are computed using  recent experimental determination of the diffractive parton density functions in Pomeron and are corrected by unitarity corrections through  gap survival probability factor. We give predictions for single as well as central diffractive ratios. These processes are sensitive to the gluon content of the Pomeron at small Bjorken-$x$ and may be particularly useful in studying the small-$x$ physics. They may also be a good place to test the different available mechanisms for quarkonium production at hadron colliders.
 
\end{abstract}


\maketitle

\section{Introduction}

Recently, it has been realized that much can be learnt about QCD from the wide variety of small-$x$ and hard diffractive processes, which are under intense study experimentally. In particular, the diffractive heavy quarkonium  production has drawn attention because their large masses provide a natural scale to guarantee the application of perturbative QCD. There are several mechanisms proposed for the quarkonium production in hadron colliders \cite{Lansberg,Kramer}, as the color singlet model, the color octet model and the color evaporation model. An important feature of these perturbative QCD models is that the cross section for quarkonium  production is expressed in terms of the product of two gluon densities at large energies. This feature is transferred to the diffractive quarkonium production, which is now  sensitive to the gluon content of the Pomeron at small-$x$ and may be particularly useful in studying the different mechanisms for quarkonium production.  We notice that at hadron colliders the heavy quarkonium production is of special significance because they have extremely clean signature through their leptonic decay modes.

Diffractive processes in hadron collisions are well described, with respect to the overall cross-sections, by Regge theory in terms of the exchange of a Pomeron  with vacuum quantum numbers \cite{Collins}. However, the nature of the Pomeron and its reaction mechanisms are not completely known. A good channel is the use of hard scattering to resolve the quark and gluon content in the Pomeron \cite{IS}. Such a parton structure is natural in a modern QCD approach to the strongly interacting Pomeron. The systematic observations of diffractive deep inelastic scattering (DDIS) at HERA have increased the knowledge about the QCD Pomeron, providing us with the diffractive distributions of singlet quarks and gluons in Pomeron as well as the diffractive structure function \cite{H1diff}. In hadronic collisions, we shall characterize an event as single diffractive if one of the colliding hadrons emits a Pomeron that scatters off the other hadron. Hard diffractive events with a large momentum transfer are also characterized by the absence of hadronic energy in certain angular regions of the final state phase space (rapidity gaps). The events fulfilling the conditions of large rapidity gaps and a highly excited hadron remnant are named single diffractive in contrast to those in which both colliding hadrons remain intact as they each emit a Pomeron (central diffractive events). In the central diffractive processes, also known as double Pomeron exchange (DPE) processes, both incoming hadrons are quasi-elastically scattered and the final states system in the center region is produced by Pomeron-Pomeron interaction. Here, we focus on the following single diffractive processes $p+p(\bar{p})\rightarrow p+J/\Psi\, [\Upsilon] +X$ and the central diffractive reactions, $p+p(\bar{p})\rightarrow p+J/\Psi\, [\Upsilon] +p(\bar{p})$.

Our starting point is the hard diffractive factorization, where the diffractive cross section is the convolution of diffractive parton distribution functions and the corresponding diffractive coefficient functions in similar way as the inclusive case. However, at high energies there are important
contributions from unitarization effects to the single-Pomeron exchange cross section. These absorptive or unitarity corrections cause the suppression of any large rapidity gap process, except elastic scattering. In the black disk limit the absorptive corrections may completely terminate those processes. This partially occurs in (anti)proton--proton collisions, where unitarity is nearly saturated at small impact parameters \cite{k3p}.  The multi-Pomeron
contributions depends, in general, on the particular
hard process and it is called survival probability factor.  At the Tevatron energy, $\sqrt s = 1.8$~TeV, the
suppression (for single diffractive processes) is of order 0.05--0.2~\cite{GLM,KMRsoft,BH,KKMR}, whereas for LHC energy, $\sqrt{s}=14$ TeV, the suppression  appears to be 0.08--0.1 ~\cite{GLM,KMRsoft,KKMR}. In the case of central diffraction, the theoretical estimates ~\cite{GLM,KMRsoft} give a survival probability factor between $0.08$ and $0.04$ for Tevatron and LHC, respectively. Therefore, these corrections are quite important for the reliability of predictions for hard diffractive processes.

The paper is organized as follows. In  next section, we present the main formulas to compute the inclusive and diffractive cross sections for heavy quarkonium hadroproduction. We also present the parameterization for the diffractive partons distribution in the Pomeron, extracted recently in DESY-HERA, and  theoretical estimations for the gap survival probability factor.  In the last section we present the  numerical results for Tevatron and perform predictions to future measurements at the LHC experiment. The compatibility with data is analyzed and the comparison with other approaches is considered.

%
%
\section{Diffractive Hadroproduction of Heavy Quarkonium}

Let us start by summarizing the formulas for inclusive and diffractive cross sections for quarkonium production in hadron colliders. For our purpose we will use the Color Evaporation Model (CEM) \cite{CEM}. The main reasons for this choice  are its simplicity and fast phenomenological implementation, which are the base for its relative success in describing high energy data (we quote Refs. \cite{Lansberg,Kramer} for a review on CEM model and competing approaches). For completeness, we notice that the CEM model employed in the analysis it is subject of some criticisms (see Sec. 1.3.3 in Ref. [1]). In this model, the cross section for a process in which partons of two hadrons, $h_1$ and $h_2$, interact to produce a heavy quarkonium state, $h_1 + h_2 \rightarrow H(nJ^{\mathrm{CP}}) + X$, is given by the cross section of open heavy-quark pair production that is summed over all spin and color states. All information on the non-perturbative transition of the $Q\bar{Q}$ pair to the heavy quarkonium $H$ of quantum numbers $J^{\mathrm{PC}}$ is contained in the factor $F_{nJ^{\mathrm{PC}}}$ that {\it a priori} depends on all quantum numbers \cite{CEM},
\begin{eqnarray}
\sigma (h_1\,h_2 \rightarrow H[nJ^{\mathrm{CP}}]\,X)= F_{nJ^{\mathrm{PC}}}\,\bar{\sigma}(h_1 \,h_2 \rightarrow Q\bar{Q}\, X)\,,
\end{eqnarray}
where $\bar{\sigma}(Q\bar{Q})$ is the total hidden cross section of open heavy-quark production calculated by integrating over the $Q\bar{Q}$ pair mass from $2m_Q$ to $2m_O$, with $m_O$ is the mass of the associated open meson. The hidden cross section can be obtained from the usual expression for the total
cross section to NLO. These hadronic cross sections in $pp$ collisions can
be written as
\begin{eqnarray}
\sigma_{pp}(\sqrt{s},m_Q^2) & = & \sum_{i,j = q, \overline q, g}
\int dx_1 \, dx_2 \,
f_i^p (x_1,\mu_F^2) \,
f_j^p(x_2,\mu_F^2)\nonumber \\ 
& \times & \widehat{\sigma}_{ij}(\sqrt{s},m_Q^2,\mu_F^2,\mu_R^2)\,,
\label{sigpp}
\end{eqnarray}
where $x_1$ and $x_2$ are the fractional momenta carried by the colliding
partons and $f_i^p$ are the proton parton densities.
The partonic cross sections are
\begin{eqnarray}
& & \widehat{\sigma}_{ij}(\sqrt{s},m_Q,\mu_F^2,\mu_R^2)  = 
\frac{\alpha_s^2(\mu_R^2)}{m_Q^2}
\left\{ f^{(0,0)}_{ij}(\rho) \right. \nonumber \\
 & + & \left. 4\pi \alpha_s(\mu_R^2) \left[f^{(1,0)}_{ij}(\rho) +
f^{(1,1)}_{ij}(\rho)\ln\bigg(\frac{\mu_F^2}{m_Q^2} \bigg) \right] 
+ {\cal O}(\alpha_s^2) \right\}
, \nonumber
\label{sigpart}
\end{eqnarray}
with $s$ the squared partonic center of mass energy, $\rho = 4m^2/s$ and
$f_{ij}^{(k,l)}$ are the scaling functions given to NLO \cite{Mangano}.
Here, we assume that the factorization scale, $\mu_F$, and
the renormalization scale, $\mu_R$, are equal, $\mu = \mu_F = \mu_R$. We also take $\mu=2m_Q$, using the quark masses $m_c=1.2$ GeV and $m_b=4.75$ GeV. These parameters provide an adequate description of open heavy-flavour production \cite{Mangano}. The invariant mass is integrated over $4m_c^2\leq \hat{s}\leq 4m_D^2$ in the charmonium case and  $4m_b^2\leq \hat{s}\leq 4m_B^2$ for $\Upsilon$ production. The factors $F_{nJ^{\mathrm{PC}}}$ are experimentally determined \cite{SV} to be $F_{11^{--}}\approx 2.5\times 10^{-2}$ for $J/\Psi$ and $F_{11^{--}}\approx 4.6\times 10^{-2}$ for $\Upsilon$. These coefficients are obtained with NLO cross sections for heavy quark production \cite{SV}.

\begin{figure}[t]
\includegraphics[scale=0.45]{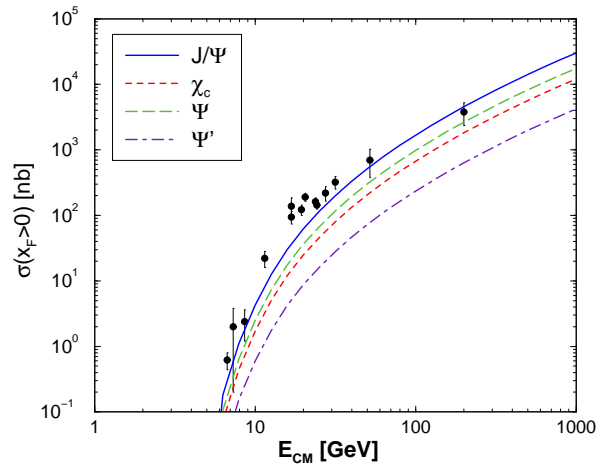}
\caption{(Color online) Total cross section ($x_F>0$) for $J/\Psi$ hadroproduction (solid line) and charmonium states computed using CEM model.  Single diffractive cross section, using single-Pomeron exchange (dot-dashed line), and available accelerator data are also shown (see text).}
\label{fig:1}
\end{figure}

In order to illustrate the success of the CEM model in describing current experimental data \cite{Lansberg}, the total cross section ($x_F>0$) for $J/\Psi$ hadroproduction is numerically computed using MRST 2004 (NLO) set of partons \cite{mrst2004nlo}. The results are presented in Fig. \ref{fig:1}, where the curves are extrapolated up to the LHC energy, $\sqrt{s}=14$ TeV. The agreement with the total cross section data  is fairly good. For energies above Tevatron, possible corrections for gluon saturation may be sizable. The results can be directly compared to previous studies in charmonium production in Ref. \cite{Gavai}. The low $x$ region is particularly relevant for $J/\Psi$ production at the LHC as well as at Tevatron. For charmonium production, the $gg$ process becomes dominant and information on the gluon distribution is of particular importance.  The CEM model is able to address the common energy behavior of the different quarkonium states. Following Ref. \cite{Gavai}, in Fig. \ref{fig:1} the energy dependence of some charmonium states are also presented. We have used the experimental values $\sigma(\psi^{\prime})/\sigma(J/\Psi)\simeq 0.14$, $\sigma(\chi_c)/\sigma(J/\psi)\simeq 0.4$ and the relative weight $\sigma(\psi)/\sigma(\psi^{\prime})\simeq 0.24$. Here, we have denoted $\psi$ by the directly produced $\mathrm{1S}$ $c\bar{c}$ state, in contrast with the experimentally  observed $J/\Psi$ (they can be originated also from radiative $\chi_c$ decays).

\begin{figure}[t]
\includegraphics[scale=0.45]{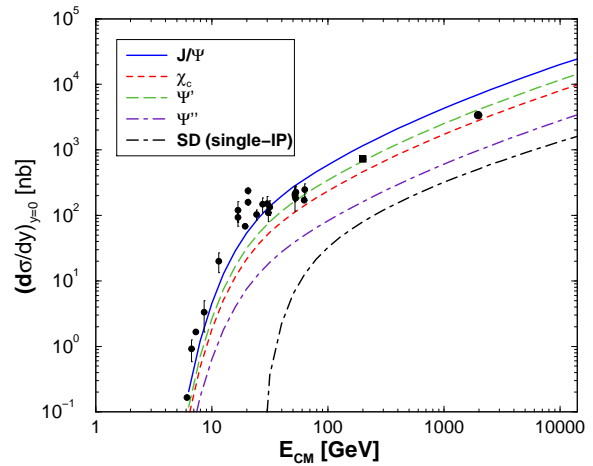}
\caption{(Color online) The differential cross section $d\sigma /dy|_{y=0}$ as a function of energy for the $J/\Psi$ production (solid line). Single diffractive cross section (dot-dashed line) and accelerator data are also shown (see text).}\label{fig:2}
\end{figure}

In Fig. \ref{fig:2}, the differential cross section $\frac{d\sigma_{pN\rightarrow J/\Psi}}{dy}|_{y=0}$ is shown as a function of center of mass energy, $\sqrt{s}$.  In this plot, the experimental data from RHIC (PHENIX point \cite{PHENIX130} at 200 GeV, from Au+Au electron measurements) and the higher energy data from Tevatron are also included. The physics involved in high energy ion--ion collisions might be very
different from the one in proton--proton collisions for the case of PHENIX data and could explain the $J/\Psi$ suppression due to a deconfined medium. The Tevatron data indicates that correction for gluon saturation may be important in bringing the theoretical curve closer to the experimental result. In particular, the typical values of Bjorken-x would be $x\sim 2 m_c/\sqrt{s}\approx 7\times 10^{-4}$ for Tevatron and $x\approx 10^{-4}$ at the LHC taken at a relatively low momentum scale $Q^2=m_{\Psi}^2=9$ GeV$^2$. In Fig. \ref{fig:3}, the differential cross section $B\times\frac{d\sigma_{pN\rightarrow \Upsilon}}{dy}|_{y=0}$  is presented as a function of the center of mass energy (solid line). In addition, the measured cross sections for the sum of the three $\Upsilon$ states ($\Upsilon+\Upsilon^{\prime}+\Upsilon^{\prime \prime}$) in the dilepton decay channel are shown \cite{Lansberg}. The agreement of the CEM model with accelerator data is very good and an extrapolation to the LHC is presented. The results give us considerable confidence in the extrapolation to the LHC energy. The predictions for direct bottomonium production in each state can be obtained by the ratios  $\sigma(\Upsilon^{\prime})/\sigma(\Upsilon)\simeq 0.36$ and $\sigma(\Upsilon^{\prime \prime})/\sigma(\Upsilon)\simeq 0.27$.

\subsection{Diffractive cross section - single--Pomeron exchange} 

For the {\it hard} diffractive
processes we will consider the Ingelman-Schlein (IS) picture \cite{IS}, where
the Pomeron structure (quark and gluon content) is probed. In the case of single diffraction, a Pomeron is emitted by one of the colliding hadrons. That hadron is detected, at least in principle, in the final state and the remaining hadron scatters off the emitted Pomeron. The diffractive cross section of a hadron--hadron collision is assumed to factorise into the total Pomeron--hadron cross
section and the  Pomeron  flux  factor  \cite{IS}.  The single
diffractive event, $h_1+h_2\rightarrow h_{1,2}+H[nJ^{\mathrm{CP}}]+X$,  may then be written as 
\begin{eqnarray}
 \label{sdexp}
& & \frac{d\sigma^{\mathrm{SD}}\,(h_i+h_j\rightarrow h_{i}+  H[nJ^{\mathrm{CP}}]+X)}{dx^{(i)}_{\pom}d|t_i|}  =  \nonumber \\
& &   F_{nJ^{\mathrm{PC}}}\times f_{{\rm\pom}/h_i}(x^{(i)}_{\pom},|t_i|)\, \bar{\sigma}\left({\pom} + h_{j}\rightarrow  Q\bar{Q}  +  X\right),
\end{eqnarray}
where the Pomeron kinematical variable $x_{\pom}$ is defined as $x_{\pom}^{(i)}=s_{\pom}^{(j)}/s_{ij}$, where  $\sqrt{s_{\pom}^{(j)}}$ is the center-of-mass energy in the Pomeron--hadron $j$ system and $\sqrt{s_{ij}}=\sqrt{s}$ the center-of-mass energy in the hadron $i$--hadron $j$ system. The momentum transfer in the  hadron $i$ vertex is denoted by $t_i$. A similar factorization can also be applied to central diffraction, where both colliding
hadrons  can in  principle  be detected in the final  state. The central quarkonium production, $h_1+h_2 \rightarrow h_1+
 H[nJ^{\mathrm{CP}}] + h_2$, is characterized by
two  quasi--elastic hadrons with  rapidity  gaps between them and the
central heavy quarkonium products. The central diffractive cross section  may then be written as,
\begin{eqnarray}
 \label{sddexp}
& & \frac{d\sigma^{\mathrm{CD}}\,(h_i+h_j\rightarrow h_i + H[nJ^{\mathrm{CP}}]   +h_j)}
{dx^{(i)}_{\pom}dx^{(j)}_{\pom}d|t_i|d|t_j|} =  F_{nJ^{\mathrm{PC}}} \times\nonumber \\
& &  f_{{\rm\pom}/i}(x^{(i)}_{\pom},|t_i|)\, f_{{\rm\pom}/j}(x^{(j)}_{\pom},|t_j|) \,\bar{\sigma}\left({\pom} + {\pom}\rightarrow  Q\bar{Q}  +  X\right).\nonumber
\end{eqnarray}

Here, we assume that one of the hadrons, say
hadron $h_1$, emits a Pomeron whose partons interact with partons of the hadron $h_2$. Thus the parton distribution  $x_1 f_{i/h_1}(x_1, \mu^2)$ in
Eq.~(\ref{sigpart}) is replaced by the convolution between a
distribution of partons in the Pomeron, $\beta f_{a/{\pom}}(\beta,
\mu^2)$, and the ``emission rate" of Pomerons by the hadron, $f_{{\pom}/h}(x_{{\pom}},t)$. The last quantity, $f_{{\pom}/h}(x_{{\pom}},t)$, is the Pomeron flux factor and its explicit formulation is described in
terms of Regge theory. Therefore, we can rewrite the parton distribution as
\begin{eqnarray}
\nonumber
\label{convol}
x_1 f_{a/h_1}(x_1, \,\mu^2) & =& \int dx_{{\pom}} \int d\beta \int dt\,
f_{{\pom}/h_1}(x_{{\pom}},\,t) \\
&\times & \beta \, f_{a/{\pom}}(\beta, \,\mu^2)\,
\delta \left(\beta-\frac{x_1}{x_{{\pom}}}\right),
\end{eqnarray}

Using  the  substitution  given  in  Eq.~(\ref{convol}), the hidden heavy flavour cross section can be obtained from Pomeron-hadron cross sections for  single  and  central diffraction processes,
\begin{eqnarray}
\label{sdxsect}
& &\frac{d\sigma\left({\pom} + h \rightarrow  Q\bar{Q}  +  X\right)}{dx_1\,dx_2}     =   \sum_{i,j=q\bar{q},g} \frac{f_{i/\pom}\left(x_1/x_{\pom}^{(1)};\,\mu^2_F\right)}{x_{\pom}^{(1)}} \nonumber \\
&\times & f_{j/h_2}(x_2,\mu^2_F)\,
\hat \sigma_{ij}(\hat{s} ,m_Q^2,\mu^2_F,\mu^2_R) +\, (1\rightleftharpoons 2)\,,
\end{eqnarray}

and

\begin{eqnarray}
\label{ddxsect}
& &\frac{d\sigma\left({\pom} + {\pom} \rightarrow  Q\bar{Q}  +  X\right)}{dx_1\,dx_2}     =  \sum_{i,j=q\bar{q},g} \frac{ f_{i/\pom}\left(x_1/x_{\pom}^{(1)};\,\mu^2_F\right)}{x_{\pom}^{(1)}}\,\nonumber \\
& & \times \frac{f_{j/\pom}\left(x_2/x_{\pom}^{(2)};\,\mu^2_F\right) }{x_{\pom}^{(2)}} \hat \sigma_{ij}(\hat{s} ,m_Q^2,\mu^2_F,\mu^2_R).\nonumber
\end{eqnarray}

In the numerical calculations, we will consider the diffractive pdf's recently obtained by the H1 Collaboration at DESY-HERA \cite{H1diff}. The Pomeron structure function has been modeled in terms of a
light flavour singlet distribution $\Sigma(z)$, consisting of $u$, $d$ and $s$
quarks and anti-quarks and a gluon distribution $g(z)$. The Pomeron carries vacuum quantum numbers, thus it is assumed that the Pomeron quark and antiquark distributions are equal and flavour independent: $q_{\pom}^f=\bar{q}_{\pom}^f=\frac{1}{2N_f}\Sigma_{\pom}$, where $\Sigma_{\pom}$ is a Pomeron singlet quark distribution and $N_f$ is the number of active flavours. Moreover, for the  Pomeron flux factor, introduced in
Eq.~(\ref{convol}), we take the experimental analysis of the diffractive structure function \cite{H1diff}, where the $x_{\pom}$ dependence is parameterized using a flux factor
motivated by Regge theory \cite{Collins},
\begin{eqnarray}
f_{\pom/p}(x_{\pom}, t) = A_{\pom} \cdot
\frac{e^{B_{\pom} t}}{x_{\pom}^{2\alpha_{\pom} (t)-1}} \ ,
\label{eq:fluxfac}
\end{eqnarray}
where  the Pomeron trajectory is assumed to be linear,
$\alpha_{\pom} (t)= \alpha_{\pom} (0) + \alpha_{\pom}^\prime t$, and the parameters
$B_{\pom}$ and $\alpha_{\pom}^\prime$ are obtained from
fits to H1 FPS data \cite{H1FPS}.

\subsection{Multiple-Pomeron exchange corrections}

In  present calculation we will consider the suppression of the hard diffractive cross section by multiple-Pomeron scattering effects. This is taken into account through a gap survival probability factor, $<\!|S|^2\!>$, which can be described in terms of screening or absorptive corrections \cite{Bj}. This suppression factor of a hard process accompanied by a rapidity gap depends not only on the probability of the initial state survive, but is sensitive to the spatial distribution of partons inside the incoming hadrons, and thus on the dynamics of the whole diffractive part of the scattering matrix. The survival factor of a large rapidity gap (LRG) in a hadronic final state is the probability of a given LRG not be filled by debris, which originate from the soft re-scattering of the spectator partons and/or from the gluon radiation emitted by partons taking part in the hard interaction. Let ${\cal A}(s,b)$ be the amplitude of the particular diffractive process of interest, considered in the impact parameter, $b$, space. Therefore, the probability that there is no extra inelastic interaction is
\begin{eqnarray}
<\!|S|^2\!>= \frac{\int d^2b\,|{\cal A}(s,b)|^2\,\exp \left[-\Omega (s,b)\right]}{\int d^2b\,|{\cal A}(s,b)|^2}\,,
\label{sdef}
\end{eqnarray}
where $\Omega$ is the opacity (or optical density) of the interaction. This quantity can be computed using a simple one-channel eikonal model or more involved multiple-channel model. The opacity $\Omega (s,b)$ reaches a maximum in the centre of proton and becomes small in the periphery. Therefore, the survival factor depends on the spatial distribution of the constituents of the relevant subprocess. For instance, the spatial $b$-distribution of single and double rapidity gap processes are assumed to be controlled by the slope $B$ of the
Pomeron-proton vertex, $\beta (t)\propto \exp(-B|t|)$, and that there is no shrinkage coming from the Pomeron amplitude associated with the LRG in hard diffractive subprocesses. In addition, $\exp (-\Omega)$ is the probability that no inelastic soft interaction in the re-scattering eikonal chain results in inelasticity of the final state at energy $s$ and impact parameter $b$.

We will consider the theoretical estimates for $<\!|S|^2\!>$ from Ref. \cite{KKMR} (labeled KMR), which considers a two-channel eikonal model and rescattering effects. The survival probability factor is computed for single, central and double diffractive processes at several energies, assuming that the spatial distribution in impact parameter space is driven by the slope $B$ of the pomeron-proton vertex. We will consider the results for single diffractive processes with $2B=5.5$ GeV$^{-2}$ (slope of the electromagnetic proton form factor) and without $N^*$ excitation, which is relevant to a forward proton spectrometer (FPS) measurement. Thus, we have $<\!|S|^2\!>_{\mathrm{KMR}}^{\mathrm{SD}}=0.15,\,[0.09]$ and  $<\!|S|^2\!>_{\mathrm{KMR}}^{\mathrm{CD}}=0.08,\,[0.04]$ for $\sqrt{s}=1.8$ TeV (Tevatron) [$\sqrt{s}=14$ TeV (LHC)]. There are similar theoretical estimates, as the GLM approach \cite{GLMrev}, which also considers a multi-channel eikonal approach. We verify that the results are consistent with each other.  We quote Ref. \cite{GLMrev} for a detailed comparison between the two approaches.

%
%

\section{Results and Discussion}
\begin{figure}[t]
\includegraphics[scale=0.47]{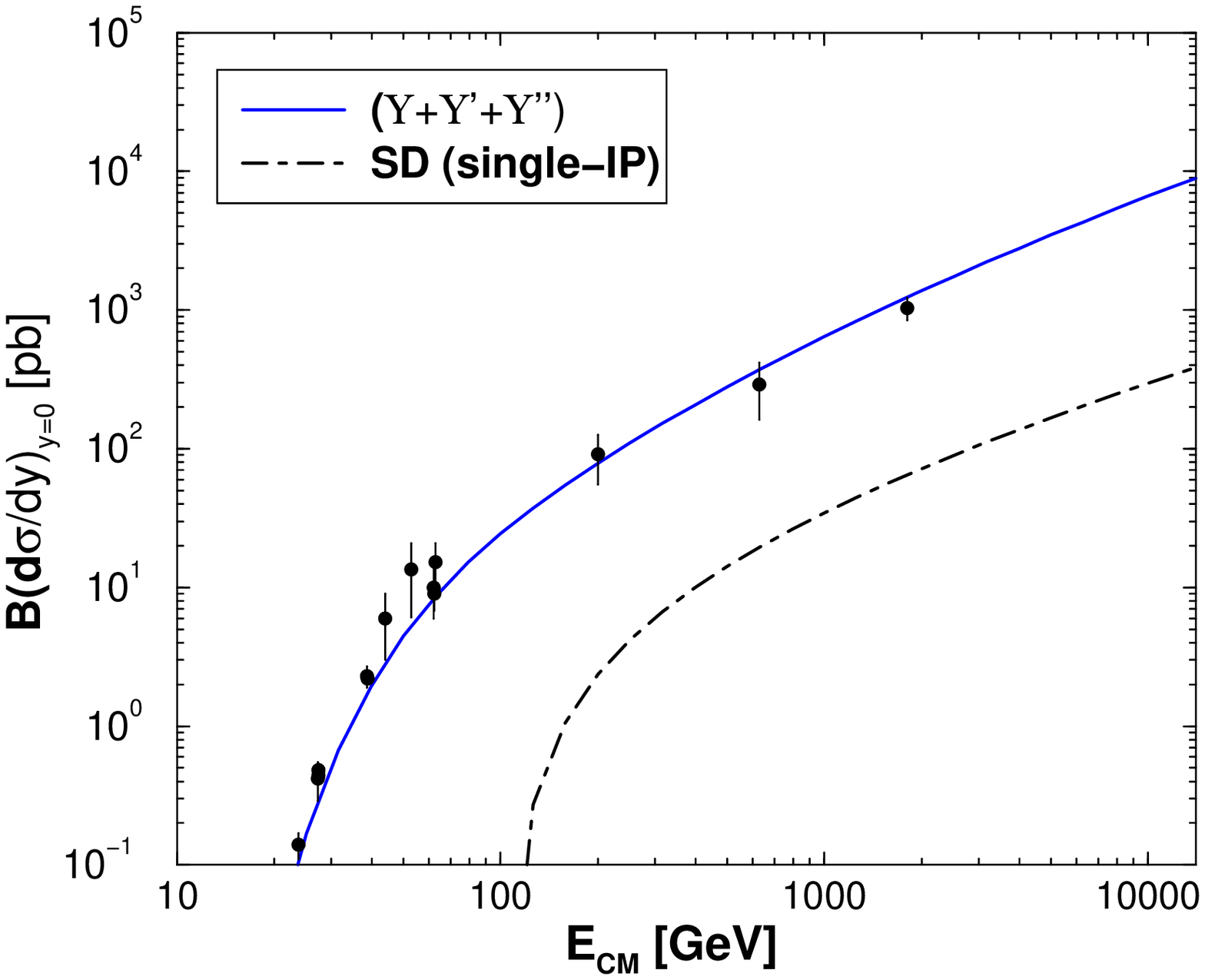}
\caption{(Color online) The differential cross section $B\times d\sigma /dy|_{y=0}$ as a function of energy for the $\Upsilon +\Upsilon^{\prime}+ \Upsilon^{\prime \prime} $ production (solid line). Single diffractive cross section (dot-dashed line) and accelerator data are also shown (see text).}
\label{fig:3}
\end{figure}

Let us now present the predictions for hard diffractive production of
heavy quarkonium.  In the numerical calculations, the new H1 parameterization for the diffractive pdf's \cite{H1diff} has been used. The `H1 2006 DPDF Fit A' is considered, with the cut $x_{\pom}<0.1$. The single-Pomeron results are presented in Figs. \ref{fig:2} and \ref{fig:3} for $J/\Psi$ and $\Upsilon$ production (dot-dashed curves), respectively. The single diffractive contribution is large, being of order 5--6 \% from the inclusive cross section.  The results corrected by multiple-Pomeron suppression factor are a factor about $1/10$ lower than the single-Pomeron ones. It should be stressed that sizable uncertainties are introduced by changing, for instance, quark masses and/or the renormalization scale. However, our purpose here is to estimate the diffractive ratios $\sigma^D/\sigma_{tot}$, which are less sensitive to a particular choice. For sake of illustration, we have $\frac{d\sigma (J/\Psi)}{dy}|_{y=0}^{\mathrm{SD}}= 85\,(566)$ nb for the energy of $\sqrt{s}=2$ TeV and $ 159 \,(1770)$ nb  for LHC energy, $\sqrt{s}=14$ TeV. For $\Upsilon$ we have $B\,\frac{d\sigma (\Upsilon)}{dy}|_{y=0}^{\mathrm{SD}}= 11\,(72)$ pb and $  35\,(386)$ pb for Tevatron and LHC, respectively. Numbers between parentheses correspond to values not corrected by survival probability gap factor.

Concerning the central difractive cross-sections, the predictions give small values. For instance, we have $\frac{d\sigma (J/\Psi)}{dy}|_{y=0}^{\mathrm{CD}}= 18$ nb  and  $B\frac{d\sigma (\Upsilon )}{dy}|_{y=0}^{\mathrm{CD}}= 0.8$ pb  at $\sqrt{s}=2$ TeV.  This gives little room to observe central diffractive $\Upsilon$ events at the Tevatron but could be promising for the LHC. The values reach $\frac{d\sigma (J/\Psi)}{dy}|_{y=0}^{\mathrm{CD}}= 45$ nb  and  $B\frac{d\sigma (\Upsilon )}{dy}|_{y=0}^{\mathrm{CD}}= 3$ pb  at  $\sqrt{s}=14$ TeV. Once again, the results corrected by multiple-Pomeron suppression factor are a factor about $1/10$ lower than the usual IS model.

Let us now compute the diffractive ratios. We have defined the single diffractive ratio as $R_{\mathrm{SD}}=\frac{d\sigma}{dy}^{\mathrm{SD}}/\frac{d\sigma}{dy}^{\mathrm{inc}} $ and the central diffractive ratio as $R_{\mathrm{CD}}=\frac{d\sigma}{dy}^{\mathrm{CD}}/\frac{d\sigma}{dy}^{\mathrm{inc}}$.   The results are summarized in Table I, where the diffractive ratios for heavy quarkonium production are presented for Tevatron and LHC energies. The multiple-Pomeron correction factors are taken from KMR model. The numbers between parentheses represent the single-Pomeron calculation. Based on these results we verify that the $J/\Psi$ and $\Upsilon$ production in single diffractive process could be observable in Tevatron and LHC, with a diffractive ratio of order of 1 \% or less. This is a similar ratio measured in $W$ and $Z$ production at the Tevatron \cite{GDMM}.  The predictions for central diffractive scattering are still not very promising, giving small ratios. However, the study of these events is worthwhile since their experimental signals are quite clear. Finally, our theoretical prediction is in good agreement with the experimental measurement of CDF \cite{CDFJPD}, which found $R_{\mathrm{SD}}^{J/\Psi}=1.45 \pm 0.25$ \%. As the main theoretical uncertainty in determining the diffractive ratio is the value for the survival factor, we consider the theoretical band $0.21$\cite{BH}--$0.15$\cite{KKMR} for Tevatron energy. Therefore, our prediction gives $R_{\mathrm{theory}}^{J/\Psi}=1.12\pm 0.19$ \%, which is consistent with the Tevatron determination.

\begin{table}[t]
\caption{\label{tab:table1} Model predictions for single and central diffractive  quarkonium  production in Tevatron and the LHC. Numbers between parentheses represent the estimates using the single Pomeron exchange.}
\begin{ruledtabular}
\begin{tabular}{lcll}
$\sqrt{s}$  & Quarkonium   & $R_{\mathrm{SD}}$ (\%) & $R_{\mathrm{CD}}$ (\%)\\
\hline
2.0 TeV &  $J/\Psi$  & $0.93\,(6.2) $   & $ 0.2\, (2.5)$\\
14 TeV &  $J/\Psi$  & $0.50\,(5.9) $  & $0.15\,(3.7)$\\
2.0 TeV &    $(\Upsilon+\Upsilon^{\prime}+\Upsilon^{\prime \prime})$  & $0.78 \,(5.2)$  & $0.06\, (0.7)$\\
14 TeV &     $(\Upsilon+\Upsilon^{\prime}+\Upsilon^{\prime \prime})$  & $0.39 \,(4.3) $   & $0.03\, (0.8)$\\
\end{tabular}
\end{ruledtabular}
\end{table}

Our calculation can be compared to available literature in diffractive heavy quarkonium production. For instance, in Ref. \cite{Yuan1} the large $p_T$ $J/\Psi$ production in hard diffractive process is computed using the color octet fragmentation mechanism and the normalized Pomeron flux \cite{Goulianos}. They found a SD $J/\Psi$ cross section at large $p_T \,(\geq 8$ GeV) of order 10 pb and a diffractive ratio $R_{\mathrm{SD}} = 0.65\pm 0.15$ \%. This is closer to our calculation for the SD ratio for Tevatron within the theoretical errors. Afterwards, in Ref. \cite{Yuan2} the color-octet mechanism combined with the two gluon exchange model(in LO approximation in QCD) for the diffractive $J/\Psi$ production is considered. Now, the SD cross section is $\sigma (p\bar{p}\rightarrow J/\Psi\,X)=66$ nb. The comparison between these calculations shows the size of the large theoretical uncertainty. A related  calculation, the $J/\Psi +\gamma $ diffractive production, appeared in Ref. \cite{Xu_Peng} based on IS model (with normalized flux \cite{Goulianos}) and factorization formalism of NRQCD for quarkonia production. They found $\sigma (p\bar{p}\rightarrow [J/\Psi +\gamma ] \,X)=$ 3.0 pb (8.5 pb) and diffactive ratio $R_{\mathrm{SD}}= 0.5\,(0.15)$ \% in central region at the Tevatron (LHC). These results are somewhat still compatible with present calculation.

Concerning central diffraction, there are some theoretical studies in literature. In Ref. \cite{Yuan3}, the DPE process, $p+p(\bar{p}) \rightarrow p+\chi_{J}+p(\bar{p})$, is calculated using two-gluon exchange model in perturbative QCD. It is found the following DPE cross sections: $\sigma(\chi_{c0})= 735$ nb and $\sigma(\chi_{b0})= 0.88$ nb. In the same work, it is found that $\frac{d\sigma (J/\Psi+\gamma)}{dydp_T}=2$ nb/GeV and $\frac{d\sigma (\Upsilon +\gamma)}{dydp_T}=0.5$ pb/GeV. Recently, in Ref. \cite{KMRSchi} the double-diffractive production of $\chi_c$ and $\chi_b$ mesons has been studied using also the Regge formalism and pQCD (including unitarity corrections). They found $d\sigma/dy|_{y=0}= 130\,(340)$ nb for $\chi_c$ production and $d\sigma/dy|_{y=0}= 0.2\,(0.6)$ nb for $\chi_b$ production for Tevatron (LHC). It was verified that the exclusive production is a factor 10 larger than the inclusive rate. Finally, the diffractive $\chi$ meson production bas been computed \cite{royon} using the Bialas-Landshoff formalism and makes use of the DPEMC Monte-Carlo simulation for small-mass diffractive production. They found $R_{\mathrm{CD}}^{\chi_c}= 6.5\,(1.6)$ \% and $R_{\mathrm{CD}}^{\chi_b}= 22\,(1.83)$ \% for Tevatron (LHC). In summary, the theoretical predictions for exclusive meson production are still quite distinct and more detailed studies are deserved.

Concerning the central diffractive quarkonia production, there are some competing processes that have been studied recently in literature. The first one is the exclusive $pp\rightarrow p+M+p$ process, with $M=J/\Psi\,(\Upsilon)$, mediated by photon exchange. In its simplest version, this process can be computed using Weizs\"{a}cker-Williams approximation where the proton is treated as a source of quasi-real photons that collide with the other proton and produce vector mesons. For instance, it has been treated in detail (going beyond the equivalent photon approach) in Refs. \cite{KMRphot,Bzdak,CZS}. Here, I quote the central scenario presented in recent study of Ref. \cite{Bzdak}: $d\sigma_{\gamma}/dy\,(pp\rightarrow pp J/\Psi)|_{y=0}\simeq 5\,(15)$ nb and  $d\sigma_{\gamma}/dy\,(pp\rightarrow pp \Upsilon)|_{y=0}\simeq 5\,(31)$ pb for Tevatron (LHC) energy. Similar results are obtained in Ref. \cite{CZS}. Another process leading to exclusive meson hadroproduction is the Pomeron-Odderon fusion mechanism, which can be investigated in the framework of Regge theory (see, for example Ref. \cite{Oddsoft}) or in the framework of $k_{\perp}$-factorization \cite{Bzdak} (the mass of the heavy vector meson supplies the hard scale in the process, allowing a treatment within perturbative QCD). For the QCD Odderon we quote results in REf. \cite{Bzdak}, where $d\sigma_{Odd}/dy\,(pp\rightarrow pp J/\Psi)|_{y=0}\simeq 1.3\,(0.9)$ nb and $d\sigma_{Odd}/dy\,(pp\rightarrow pp \Upsilon)|_{y=0}\simeq 4\,(5)$ pb for Tevatron (LHC) energy. Let us compare these values with the results for central Pomeron-Pomeron production presented here: $d\sigma_{\mathrm{CD}}/dy\,(J/\Psi)|_{y=0}\simeq 18\,(45)$ nb and  $d\sigma_{\mathrm{CD}}/dy\,(\Upsilon)|_{y=0}\simeq 32\,(120)$ pb for Tevatron (LHC) energy, which include also the absobtive corrections. This indicates that the central diffractive contribution is substantially larger than photon and/or Odderon exchange. However, the different competing processes could be identified by using additional experimental cuts as the outgoing proton momenta distibution.

Finally, some comments are in order on the limitations of the present study. The theoretical uncertainties in both, quarkonium production and diffraction, are large and changes in the quarkonium models are
likely to be compensated by changes in the diffractive part, the value
of the heavy quark mass and of the factorization scale (and vice versa). A more detailed study should be done to clarify these issues. For example, in order to demonstrate the sensitivity to the diffractive gluon,
it would have been nice to propagate the uncertainty band of the H1 Fit A
gluons or to compare with the Fit B gluons. Also the variation
of the scales and heavy quark mass would have been easily possible and
interesting for the absolute cross sections in Figs. 1-3 but also for
the diffractive ratios to see how much these variations contribute to
the theoretical error which is dominated by the uncertainty of the
gap survival factor.  Reliable estimates of the uncertainties are
of particular importance because the existing
theoretical predictions for exclusive meson production are quite distinct.

In summary, we have presented predictions for diffractive heavy quarkonium production at the Tevatron and the LHC. We use Regge factorization, corrected by unitarity corrections modeled by a gap survival probability factor (correction for multiple-Pomeron exchange). The perturbative formalism for meson hadroproduction is based on the CEM model, which is quite successful in describing experimental results for inclusive production. For the Pomeron structure function, recent  H1 diffractive parton density functions have been used. The results are directly dependent on the quark singlet and gluon content of the Pomeron.  We estimate the multiple interaction corrections taking the theoretical prediction a  multi-channel model (KMR), where the gap factor decreases on energy. That is,  $<\!|S|^2\!>\simeq 15$ \% for Tevatron energies going down to $<\!|S|^2\!>\simeq 9$ \% at LHC energy (for single diffractive process). We found that at the Tevatron single and central diffractive $J/\Psi$ and $\Upsilon$  production is observable with a single diffractive ratio $R^{\mathrm{SD}}(\mathrm{Tevatron})$ that is between  1 \% (charmonium) and 0.5 \% (bottomonium), with  lower values at the  LHC. In particular, we predict  $R_{\mathrm{theory}}^{J/\Psi}=1.12 \pm 0.19 $ \%, which is in agreement with Tevatron measurements. The central diffractive cross sections for quarkonium  production are still feasible to be measured, despite the very small diffractive ratios. In this case, the theoretical model dependence is still very large and detailed studies are deserved.


\section*{Acknowledgments}

This work was supported by CNPq, Brazil. The author thanks the hospitality of the Center for High Energy Physics (CHEP) at Kyungpook National University and the pleasant atmosphere of the Lepton Photon 2007 Conference in Daegu, Republic of Korea,  where this work was accomplished.


%
%


\end{document}